\documentclass[final,3p,onecolumn]{elsarticle}

\usepackage{amssymb}
\usepackage{xcolor}
\usepackage{graphicx}
\usepackage{caption}

\journal{International Journal of Modern Physics A}

\begin{document}

\begin{frontmatter}

\title{Pseudo-Hermitian Dirac operator on the torus for massless fermions under the action of external fields}

\author[gazi]{\"{O}. Ye\c{s}ilta\c{s}\corref{cor1}}
\address[gazi]{Department of Physics, Faculty of Science, Gazi University, 06500 Ankara, Turkey}
\ead{yesiltas@gazi.edu.tr}
\cortext[cor1]{yesiltas@gazi.edu.tr}

\author[ufca]{J. Furtado}

\address[ufca]{Universidade Federal do Cariri(UFCA), Av. Tenente Raimundo Rocha, \\ Cidade Universit\'{a}ria, Juazeiro do Norte, Cear\'{a}, CEP 63048-080, Brasil}

\date{\today}

\begin{abstract}
The Dirac equation in $(2+1)$ dimensions  on the toroidal surface is studied for a massless fermion particle under the action of external fields. Using the covariant approach based in general relativity, the Dirac operator stemming from a metric related to the strain tensor is discussed within the Pseudo-Hermitian operator theory. Furthermore, analytical solutions are obtained for two cases, namely, constant and position-dependent Fermi velocity.
\end{abstract}


\end{frontmatter}


\section{Introduction}

When gravity meets quantum theory, there is mutual incompatibility between general relativity and quantum mechanics. In this sense, quantum gravity is one of the most popular and essential topics which is being aimed to become a working physical theory. Besides its complexities, there are fundamental physics problems involving the interaction between an atom and the gravitational field that can be examined with Dirac equation in a curved spacetime where the spacetime curvature can change the phase of the wave function and thus the curvature effect is restricted by the atomic spectrum. In this context, among the fundamental quantum field investigations one can highlight the studies such as the perturbations of the energy levels of an atom in a gravitational field by Parker \cite{01}, particle creation \cite{02}, spinning objects in a curved spacetime \cite{03} and transformation techniques for the curved Dirac equation in gravity into a Dirac equation in flat spacetime with the exact solutions and scattering analysis \cite{04, 05}.



In the context of low energy physics, the importance of technological advances can bring a new sight into Dirac's theory and its symmetries. The growing interest in two-dimensional materials such graphene is drawing more attention to $(2+1)$ dimensional physics \cite{0}. In this point of view, the unique properties of graphene, which is an atomic honeycomb lattice made of carbon atoms, has opened a way in a wide spectrum of applications ranging from electronics to optics and nanotechnology since its discovery \cite{N}. A single layer graphene presents no gap in the conductance band so that an electron in its surface is governed by a linear dispersion relation, behaving as a relativistic massless particle described by Dirac equation \cite{KT}. 

The topology of graphene requires $(2+1)$ dimensional Dirac equation  which allows the study curvature effects in the lattice \cite{PR} and it is pointed out that curvature of the graphene changes the electron density of the states \cite{1}. Inherently, different geometries can bring different curvature effects in the lattice and curvature can alter the electron density. Therefore, the possibility of constructing new electronic devices based on curved graphene structures has motivated the study of graphene in several curved surfaces, such as M\"{o}bius-strip \cite{GUO}, ripples \cite{JU}, corrugated surfaces \cite{ATA}, catenoid \cite{JOB}, among others. The intrinsic curvature and strain effects are discussed through $(2+1)$ dimensional Dirac equation in \cite{io}. Graphene nanoribbons are discussed using the long-wave approximation in \cite{2}. Electronic structure of a helicoidal graphene and the scattering states can be found in \cite{heli}. It is important to highlight here that elegant methods in quantum mechanics, such as supersymmetry, can be used to describe curvature effects on carbon nanostructures \cite{jak1, jak2}, and in particular exact solutions of Dirac equation \cite{cast, Bagchi}. In \cite{cast}, the authors have studied the behaviour of a Dirac electron in graphene under the action of a magnetic field orthogonal to the layer by using supersymmetric quantum mechanics, while in \cite{Bagchi} the authors have investigated the most general form of the one-dimensional Dirac Hamiltonian in the presence of scalar and pseudoscalar potentials in the framework of supersymmetric quantum mechanics.

An important geometry intensively studied in the last years is the torus surface. Curvature effects plays an important role in toroidal geometry \cite{t1}. Toroidal carbon nanotubes, also known as carbon nanotori, appears in nanoelectronics, quantum computing, and  biosensors  \cite{t2, t3}. Considering an electron governed by the Schr\"{o}dinger equation, the curvature-induced bound-state eigenvalues and eigenfunctions were calculated for a particle constrained to move on a torus surface in \cite{ENC}. Under these same considerations the action of external fields was addressed in \cite{EUC}. A charged spin $1/2$ particle, governed by Pauli equation, moving along a toroidal surface was studied in \cite{AGM}. Exact solutions of $(2+1)$ Dirac equation on the torus were first obtained in \cite{OZL} using supersymmetric quantum mechanics for two cases, constant and position-dependent Fermi velocity. As far as we know, the non-constant Fermi velocity was first studied in \cite{voz}, and posterior works such as \cite{mustafa, oliva, Ghosh1} presented new insights on the topic. The consideration of position-dependent Fermi velocity could be thought as an effective way of treating the lattice strain. As a natural continuation of the work \cite{OZL}, in this paper we study the Dirac equation in $(2+1)$ dimensions on the toroidal surface for a massless fermion particle under the action of external fields. Using the covariant approach based in general relativity, the Dirac operator stemming from a metric related to the strain tensor is discussed within the Pseudo-Hermitian operator theory. Furthermore, analytical solutions are obtained for two cases, namely, constant and position-dependent Fermi velocity.


This paper is organized as follows: In section 2 we discuss the Dirac equation on the torus and we decouple the left and right sectors of the spinor in order to obtain two Klein-Gordon-like equations for the system for two cases, namely, constant and position-dependent Fermi velocity. In section 3 we present the pseudo-Hermitian operators as well as the pseudo-supersymmetry for the system in both cases. In section 4, the point canonical transformation are used in order to obtain the solutions. The conclusions are given in section 5.



\section{Dirac equation on the torus}
\begin{figure}[ht!]
    \centering
    \includegraphics[scale=0.5]{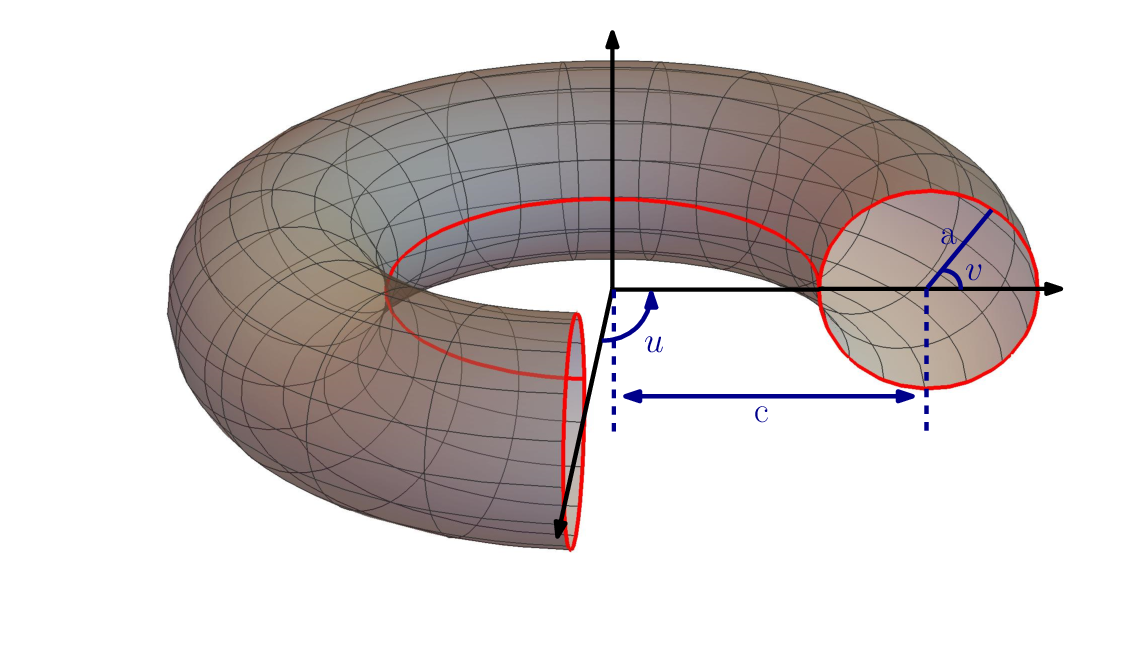}
    \caption{Torus section}
    \label{fig1}
\end{figure}

Condensed matter physics has been witnessed an important evolution in the study of massless fermions on the surface of graphene which devotes the interest of the community of both condensed matter and relativity theorists. The massless Dirac equation, written as 
\begin{equation}\label{01}
  i \gamma^{\mu}\partial_{\mu}\psi=E\psi,
\end{equation}
describes the dynamics of a low energy electron in a flat surface of graphene. Here $\gamma^{\mu}$ are Dirac matrices. Moreover, the Dirac equation can be generalized to the curved spacetime in terms of covariant derivatives, vierbein fields and spin connection as \cite{2}
\begin{equation}\label{1}
  [i\gamma^{\mu}(\partial_{\mu}-\Gamma_{\mu}+ieA_{\mu}) ]\Psi=0,
\end{equation}
where $\Gamma_{\mu}$ is the spin connection, $A_{\mu}$ is the gauge field and 
\begin{equation}
 \Psi=\left(\begin{array}{cc}
      \Psi_1 \\
      \Psi_2
 \end{array}\right)   
\end{equation}
is the spinor which includes electron's wave-functions near the Dirac point. The Dirac matrices $\gamma^{\mu}$ in curved spacetime satisfy the Clifford algebra, so that,
\begin{equation}\label{2}
  \{\gamma^{\mu}, \gamma^{\nu}\}=2g^{\mu \nu},
\end{equation}
and
\begin{equation}\label{gt}
  \gamma^{\mu}(x)=e^{\mu}_{i}\bar{\gamma}^{i}.
\end{equation}
Here $g^{\mu\nu}$ is the metric tensor and the tetrad(vierbein) frames field is defined as
\begin{equation}\label{3}
    g_{\mu \nu}=e^{a}_{\mu}e^{b}_{\nu}\eta_{ab}
\end{equation}
where $\eta_{ab}=diag(1,-1,-1)$. The Greek and Roman letters correspond to global and local indices respectively. Additionally, the metric  for the torus surface is given by
\begin{equation}\label{4}
    ds^{2}=(dx^{0})^{2}-a^{2}dv^{2}-(c+a \cos v)^{2}du^{2}.
\end{equation}
In the metric given above,  the inner radius of the torus is $c$, the outer radius is shown by $a$, $c\neq a$ and $u, v \in [0, 2\pi )$. Besides, the angle going round the big sweep of the torus from $0$ to $2\pi$ is $u$ and the angle going around the little waist of the torus through the same interval is $v$, as you can see in figure (\ref{fig1}). In (\ref{1}), we can use the spin connection formula which is

\begin{equation}\label{5}
    \Gamma_{\mu}=\frac{1}{2}S^{ab}e_{a}^{\nu}g_{\rho \nu}D_{\mu}e_{b}^{\rho}
\end{equation}
where $x^{1}=v=x, ~~ x^{2}=u$. Moreover, the $S^{ab}$ spin matrix and covariant derivatives on zweibeins are
\begin{eqnarray}
  S^{ab} &=& \frac{1}{4}[\bar{\gamma}^{a}, \bar{\gamma}^{b}], \\
  D_{\nu} e^{a}_{\mu} &=& \partial _{\nu}e_{\mu}^{a} + e_{\nu}^{b}\omega^{a}_{b \mu}-e_{\lambda}^{a}\Gamma^{\lambda}_{\nu \mu}.
\end{eqnarray}
In the tetrad formalism \cite{collas}, a set of $n$ independent vector fields are defined as
\begin{equation}\label{col}
  e_{a}=e^{\mu}_{a} \partial_{\mu},~~~~e^{a}=e_{\mu}^{a}dx^{\mu},
\end{equation}
where  a vierbein is identified as the coefficients $e_{a}^{\mu}$. In \cite{1}, the Christoffel symbols $\Gamma^{\lambda}_{\mu\sigma}$ were given in terms of the variable $R(x)=c+a \cos x$. Then, the nonvanishing components of the Christoffel symbols are:
\begin{equation}\label{6}
    \Gamma^{2}_{12}=-\frac{a\sin x}{R(x)},~~ \Gamma^{1}_{22}=\frac{1}{a}R(x)\sin x
\end{equation}
hence $\Gamma_2$ can be obtained as,
\begin{equation}\label{7}
    \Gamma_2=\frac{a}{2}R(x)\sin x \gamma_{1}\gamma_2.
\end{equation}
We also note that the vierbeins read as
\begin{equation}\label{v}
 e^{i}_{\mu}= \left(
    \begin{array}{ccc}
      1 & 0 & 0 \\
      0 & a & 0 \\
      0 & 0 & R(x) \\
    \end{array}
  \right).
\end{equation}
Using (\ref{col}), (\ref{7}) and (\ref{1}), one can obtain
\begin{equation}\label{d1}
 \left[ \frac{i}{V_{F}}\bar{\gamma}^{0} \frac{d}{dt}+\frac{i}{a}\bar{\gamma}^{1}\left(\frac{d}{dx}-\frac{a^{2}}{2}\sin x+ieA_{x}(x)\right)+\bar{\gamma}^{2}\left(\frac{i}{R(x)} \frac{d}{du}-\frac{e}{R(x)}A_{u}(x)\right)\right]\Psi=0,
\end{equation}
where the Dirac matrices in flat spacetime $\bar{\gamma}^{i}$ are written in terms of Pauli matrices $\sigma^{i}$ as
\begin{equation}\label{d11}
  \bar{\gamma}^{0}=\sigma^{3}, \bar{\gamma}^{1}=-i\sigma^{2}, \bar{\gamma}^{2}=-i\sigma^{1}.
\end{equation}
We also note that $V_F$ stands for the Fermi velocity. Thus,  we get
\begin{equation}\label{d2}
  H_{D}\Psi(X)=i\frac{d}{dt}\Psi(X)
\end{equation}
where
\begin{equation}\label{d3}
  H_{D}=-\frac{1}{a}\sigma^{1}\left(\frac{d}{dx}-\frac{a^{2}}{2}\sin x+ieA_{x}(x)\right)+\sigma^{2}\left(-\frac{1}{R(x)} \frac{d}{du}-ia\frac{e}{R(x)}A_{u}(x)\right),
\end{equation}
and $X=x^{\mu}$.
\subsection{Hermiticity}
Next we look at the Hermiticity of $H_D$ by noting that $A_{x}(x)$ and $A_{u}(x)$ are real functions. For the stationary states of the Dirac spinor $\Psi(X)=\exp(iEt)\Psi(x,u)$, we have
\begin{equation}\label{d4}
 H_D \Psi(x,u)=\left(\frac{E}{V_{F}} \right) \Psi(x,u).
\end{equation}
It can be seen that the operator $H_{D}$ in (\ref{d3}) is non-Hermitian, i.e. $H_D  \neq H^{\dag}_D$, and the matrix representation of $H_D$ can be given by
\begin{equation}\label{mr}
 H_{D}= \left(
    \begin{array}{cc}
      0 & -\frac{1}{a}\frac{d}{dx}+\frac{i}{R(x)}\frac{d}{du}+W_1(x)-iW_2(x) \\
     -\frac{1}{a}\frac{d}{dx}-\frac{i}{R(x)}\frac{d}{du}+W_1(x)+iW_2(x) & 0 \\
    \end{array}
  \right)
\end{equation}
where
\begin{eqnarray}
  W_1(x) &=& \frac{a}{2}\sin x-\frac{ie}{a}A_{x}(x) \\
  W_2(x) &=& -ie\frac{a}{R(x)}A_{u}(x),
\end{eqnarray}
and $W_{1}(x) \neq W^{*}_{1}(x), W_{2}(x)\neq W^{*}_{2}(x)$, and $^{*}$ stands for the complex conjugation. In case of imaginary $A_{x}(x)$ or $A_{x}=0$, $H_{D}$ becomes Hermitian. In the next, we will look at the properties of  $H_{D}$  in more detail. \\
\\
\textbf{Case 1: Real vector potential components and constant Fermi velocity}\\\\
We can define the two component spinors
\begin{equation}\label{d6}
  \Psi(x,u)=\exp(iku)\left(\begin{array}{cc}
       \Psi_{1}(x)  \\
       \Psi_2(x) 
  \end{array}\right).
\end{equation}
Hence, we obtain
\begin{eqnarray}\label{d06}
  -\Psi^{''}_{1}(x)+\sigma(x)\Psi^{'}_{1}(x)+\rho^{+}(x)\Psi_{1}(x) &=& \epsilon^{2}a^{2} \Psi_1(x)   \\
  -\Psi^{''}_{2}(x)+\sigma(x)\Psi^{'}_{2}(x)+\rho^{-}(x)\Psi_{2}(x) &=&  \epsilon^{2} a^{2}\Psi_2(x), \label{d60}
\end{eqnarray}
where $\epsilon=\frac{E}{V_{F}}$ and
\begin{eqnarray}
  \sigma(x) &=& a^{2}\sin x-2ieA_{x}(x)  \\
  \rho^{+}(x) &=& -ieA^{'}_{x}(x)+\left(eA_{x}(x)+i\frac{a^{2}}{2}\sin x\right)^{2}+\frac{a^{2}}{2}\cos x+\frac{(ka+a^{2}e A_{u}(x))^{2}}{R(x)^{2}}  \nonumber   \\ \nonumber
  &  & + \frac{aeA^{'}_{u}(x)}{R(x)} -\frac{kaR'(x)}{R(x)^{2}}-\frac{a^{2}eA_{u}(x)R'(x)}{R(x)^{2}}  \\  \nonumber
\rho^{-}(x)  & = & \rho^{+}(x)|_{(k\rightarrow -k, ~~ A_{u}(x)\rightarrow -A_{u}(x))}.
\end{eqnarray}

\textbf{Case 2: Real vector potential components and position dependent Fermi velocity}\\\\
Next, we look at the same Hamiltonian in (\ref{d3}) but $V_F$ is taken as position-dependent function. It is interesting to consider a position-dependent Fermi velocity, i.e., $V_F=V_F(x)$, since such dependence is an effective way of considering effects of strain. The dependence of the Fermi velocity as a function only of $x$ lies in the symmetry of the torus on the angular variable, so that no dependence of $u$ is expected.
  Using (\ref{d6}),
\begin{equation}\label{dd5}
    \left[\frac{i}{V_{F}(x)}\sigma^{0} \frac{d}{dt}+\frac{i}{a}\sigma^{1}\left(\frac{d}{dx}-\frac{a^{2}}{2}\sin x+ieA_{x}(x)\right)+\sigma^{2}\left(\frac{i}{R(x)} \frac{d}{du}-a\frac{e}{R(x)}A_{u}(x)\right)\right]\Psi=0,
\end{equation}
which leads to a couple of differential equations,
\begin{eqnarray}
  \frac{E_n}{V_F(x)}\Psi_1(x) &=& \left(\frac{1}{a}\frac{\partial }{\partial x}+\frac{a^{2}}{2}\sin x-ieA_x(x)-\frac{k}{R(x)}-\frac{aeA_u(x)}{R(x)}\right)\Psi_2(x) \\
 \frac{E_n}{V_F(x)}\Psi_2(x) &=& \left(\frac{1}{a}\frac{\partial }{\partial x}+\frac{a^{2}}{2}\sin x-ieA_x(x)-\frac{k}{R(x)}-\frac{aeA_u(x)}{R(x)}\right)\Psi_1(x).
\end{eqnarray}
Hence, a couple of second order differential equations can be obtained as
\begin{eqnarray}\label{ddd5}
   \nonumber-V_{F}^{2}(x)\Psi''_{1}(x)&+&[-V_{F}V'_{F}+(a^{2}\sin x-2ieA_{x}(x))]V^{2}_{F}\Psi'_1(x)+\\
   &&+[F^{+}(x)V^{2}_{F}+G^{+}(x)V_F(x)V'_F(x)]\Psi_{1}(x) =a^{2}E^{2}\Psi_1(x)   \\
 \nonumber-V_{F}^{2}(x)\Psi''_{2}(x)&+&[-V_{F}V'_{F}+(a^{2}\sin x-2ieA_{x}(x))]V^{2}_{F}\Psi'_2(x)+\\
 &&+[F^{-}(x)V^{2}_{F}+G^{-}(x)V_F(x)V'_F(x)]\Psi_{2}(x)   =  a^{2}E^{2}\Psi_2(x) \label{dddd5}
\end{eqnarray}
where
\begin{eqnarray}
  F^{+}(x) &=& e^{2}A^{2}_{x}(x)+\frac{a^{2}}{2}\cos x-\frac{a^{4}}{4}\sin x+\frac{(k a+a^{2}e A_{u}(x))^{2}}{R(x)^{2}}
  +ie(a^{2}A_x \sin x-A'_{x})+\frac{a^{2}}{R}eA'_{u}(x) \nonumber \\ \nonumber & & +\frac{a R'(x)((k+a eA_{u}(x)))}{R(x)^{2}} \nonumber  \\
  G^{+}(x) &=& -ieA_{x}(x)+\frac{a k}{R(x)}+\frac{a^{2}eA_{u}(x)}{R(x)}+\frac{a^{2}}{2}\sin x \nonumber
\end{eqnarray}
and
\begin{equation}\label{fg}
  F^{-}(x)=F^{+}(x)|_{k\rightarrow -k, A_{u}(x)\rightarrow -A_{u}(x)},~~~~G^{-}(x)=G^{+}(x)|_{k\rightarrow -k, A_{u}(x)\rightarrow -A_{u}(x)}.
\end{equation}
Let us discuss now the pseudo-Hermitian operators in the present context.

\section{Pseudo-Hermitian Operators}
\subsection{Hilbert Space}
Let $\mathfrak{H}$ be the Hilbert space and $\mathcal{O}:\mathfrak{H}_{+}\rightarrow \mathfrak{H}_{-}$ a linear operator. A class of non-Hermitian operators is the pseudo-Hermitian operators \cite{am} satisfying the similarity transformation given as
\begin{equation}\label{d7}
  \mathcal{O}^{\dag}=\eta \mathcal{O} \eta^{-1}
\end{equation}
where $\eta$ is the invertible and linear operator. In the basic properties of pseudo-Hermitian operators, one can remember that the eigenvalues of $\mathcal{O}$ are either real or complex conjugate pairs and the operator commutes with an invertible antilinear operator. If the operator is pseudo-Hermitian, there are infinite number of $\eta$ which satisfy (\ref{d7}), $\eta_{\pm}:\mathfrak{H}_{\pm}\rightarrow \mathfrak{H}_{\mp}$. Moreover, the pseudo-adjoint of $\mathcal{O}$ is,  $\mathcal{O}^{\sharp}: \mathfrak{H}_{-}\rightarrow \mathfrak{H}_{+}$. And it is given by
\begin{equation}\label{psa}
  \mathcal{O}^{\sharp}=\eta^{-1}_{+}\mathcal{O}\eta_{-}
\end{equation}
If $\mathfrak{H}_{+}=\mathfrak{H}_{-}$ and $\eta_{-}=\eta_{+}=\eta$ and a quantum Hamilton operator $H$ is pseudo-Hermitian, i.e.
\begin{equation}\label{herr}
  H^{\dag}=\eta H \eta^{-1}.
\end{equation}
This operator $H$ can be also factorible within the first order differential operators $L_{1}, L_{2}$:
\begin{equation}\label{d8}
  H=L_{1}L_{2},
\end{equation}
and $H_{S}$ is the partner operator which is given by
\begin{equation}\label{d08}
  H_{S}=L_{2}L_{_{1}},
\end{equation}
and we note that the adjoint of $H_{S}$ is $H^{\dag}_{S}$. We can link $H$ to $H^{\dag}_{S}$ using the intertwining relation given below
\begin{equation}\label{d9}
  \bar{\eta} H= H^{\dag}_{S}\bar{ \eta}^{-1},
\end{equation}
where $\bar{\eta}=\eta_{2}\eta_{1}$. One can look at the proof of (\ref{d9}) in \cite{roy}. The operators $H$, $H_{S}$, $H^{\dag}_{p}$ satisfy the relationships given below \cite{roy},
\begin{equation}\label{d10}
  \eta_{1} H=H_{S} \eta_{1},
\end{equation}
\begin{equation}\label{d11}
  \eta_2 H_p=H^{\dag}_{S} \eta_2
\end{equation}
On the other hand, we may give the intertwining operator relations as below
\begin{equation}\label{d12}
  L_{1}H=H_{S}L_{1},~~~~L_{2}H_{S}=H L_{2}.
\end{equation}
We note that $L_{2}=L^{\sharp}_{1}$.
\subsection{Pseudo-supersymmetry for the torus-Dirac system}
\subsubsection{Constant Fermi velocity}
Let $\mathcal{H}_{S}$ be the Hamiltonian linked to the system given in (\ref{d06}) and (\ref{d60}):
\begin{equation}\label{d13}
  \mathcal{H}_{S}=-\frac{d^{2}}{dx^{2}}+\sigma(x)\frac{d}{dx}+\rho^{(+,-)}(x).
\end{equation}
The Hermitian counterpart of (\ref{d13}) can be found by
\begin{equation}\label{d14}
  \mathcal{H}^{\dag}_{S}\eta_{2}=\eta_{2} \mathcal{H}_{S},
\end{equation}
where
\begin{equation}\label{d15}
  \eta_{2}(x)=\frac{d}{dx}+A(x),
\end{equation}
here $A(x)$ is an unknown function which will be found using (\ref{d14}). We can show the findings in order to satisfy (\ref{d14}) as
\begin{eqnarray}
  A_{x}(x) &=&  -\frac{ia^{2}\sin x}{2e} \\
  A(x) &=& C_1+ \frac{a^{4}x}{4}-\frac{a^{2}}{2}\sin x-\frac{a^{4}\sin 2x}{8}.
\end{eqnarray}
Then, the Hermitian partner of $\mathcal{H}_{S}$ can be found as,
\begin{equation}\label{d16}
  \mathcal{H}_{S}^{\dag}=-\frac{d^{2}}{dx^{2}}+\frac{(ak+a^{2}eA_{u}(x))^{2}}{R(x)^{2}}+\frac{aeA'_{u}(x)}{R(x)}-
  \frac{akR'(x)}{R(x)^{2}}-\frac{a^{2}eA_{u}(x)R'(x)}{R(x)^{2}}.
\end{equation}
One may also be interested in the exact solutions of (\ref{d16}). Substituting $A_{u}(x)$ in the potential function of (\ref{d16}) as
\begin{equation}\label{d17}
  A_{u}(x)=C_{2}R(x)^{2}+C_{3}
\end{equation}
\begin{eqnarray}\label{d18}
  V_{1}(x) &=& \frac{(ak+a^{2}eA_{u}(x))^{2}}{R(x)^{2}}+\frac{aeA'_{u}(x)}{R(x)}-
  \frac{akR'(x)}{R(x)^{2}}-\frac{a^{2}eA_{u}(x)R'(x)}{R(x)^{2}} \\
   &=& a^{4}C^{2}_{2}e^{2}R(x)^{2}+2aC_{2}eR'(x)-a^{2}C_{2}eR'(x).
\end{eqnarray}
Here, $C_3=-\frac{k}{a e}$ is used to get $V_{1}(x)$ in terms of $R(x)$ and $R'(x)$ above. Let us recall the potential model which is known as Mathieu potential in the literature \cite{dong}
\begin{equation}\label{d19}
  U(x)=B^{2}\sin^{2}bx-2B b (2c+1)\cos bx+\nu b^{2},~~B>0,~~b>0.
\end{equation}
Let us express $V_1(x)$ in the form of $U(x)$:
\begin{equation}\label{d20}
 V_1(x)=a^{4}(a^{2}+c^{2})C^{2}_{2}e^{2}+2a^{5}cC^{2}_{2}e^{2}\cos x+eC_2 a^{2}(a-2)\sin x
  -a^{6}C^{2}_{2}e^{2}\sin^{2}x.
\end{equation}
The term $\sin x$ doesn't match with the model in \cite{dong}. We will find the exact solutions of (\ref{d20}) in the next section. Now we continue with the pseudo-Hermiticity properties of the problem. If we turn back to $\mathcal{H}^{\dag}_{S}$, let us factorise it and then, we obtain $\mathcal{H}_{1}$ which is the supersymmetric partner Hamiltonian of $\mathcal{H}_{S}$. Hence,
\begin{equation}\label{d21}
  \mathcal{H}^{\dag}_{S}=A^{\dag}A,~~A=\frac{d}{dx}+W(x),
\end{equation}
where
\begin{equation}\label{d22}
  W(x)=-\frac{i\sqrt{a-1}}{a}\sin x+\frac{i(a-2)}{2a},
\end{equation}
where $W(x)$ is the superpotential and the constants $C_2$ and $c$ shall satisfy the following conditions
\begin{eqnarray}\label{C2}
  C_2 &=& \frac{\sqrt{a-1}}{a^{4}e} \\
  c &=& \frac{1}{2}\frac{a^{2}}{\sqrt{1-a}}.
\end{eqnarray}
Since $c$ is the outer radius of the torus, it must be real number and this brings a constraint for the inner radius $a<1$. Hence, the symmetry leads to a condition on the torus parameters. We note that $\mathcal{H}_{1}$ can also be obtained using $\mathcal{H}_{1}=AA^{\dag}$:
\begin{equation}\label{d23}
  \mathcal{H}_{1}=-\frac{d^{2}}{dx^{^{2}}}+V_{1}(x)
\end{equation}
where $V_{1}$ can be obtained as
\begin{equation}\label{d24}
  V_{1}(x)= \frac{a-1}{a^{2}}\cos^{2}x+\frac{a-2}{a^{2}}\sqrt{a-1}\sin x-\frac{i\sqrt{a-1}}{a}\cos x-\frac{1}{4},
\end{equation}
while $V(x)$ was obtained as
\begin{equation}\label{d25}
  V(x)=\frac{a-1}{a^{2}}\cos^{2}x+\frac{a-2}{a^{2}}\sqrt{a-1}\sin x+\frac{i\sqrt{a-1}}{a}\cos x-\frac{1}{4}.
\end{equation}
Now let us obtain $\eta_1$ operator using
\begin{equation}\label{d26}
  \mathcal{H}_{1}=\eta_{1}^{-1}\mathcal{H}_{S} \eta_{1}.
\end{equation}
And we get,
\begin{equation}\label{d27}
  \eta_1(x)=\frac{i(2-a)}{2a}+\frac{i\sqrt{a-1}}{a}\sin x.
\end{equation}
We have constructed the pseudosupersymmetry of the system in (\ref{d06}) and (\ref{d60}). Final effort shall be given in order to express $\mathcal{H}_{1}$ in the form of (\ref{d13}).
\subsubsection{position-dependent Fermi velocity}
For the system given in (\ref{ddd5}) and (\ref{dddd5}), the intertwining operator is given by
\begin{equation}\label{d28}
  \eta_2(x)=\frac{d}{dx}+\frac{a^{4}}{16}+C_2+\frac{3}{4}a^{2}\sin x-\frac{a^{4}}{32}\sin 2x.
\end{equation}
By the way, we can mention the Sturm-Liouville equation in (\ref{ddd5}) and search for  physical model. For the sake of simplicity, we will discuss the partner Hamiltonian representations afterwards. Using (\ref{d28}), the Hermitian counterpart of the Hamiltonian operator corresponding to (\ref{d06}) becomes
\begin{equation}\label{d29}
 \mathcal{H}^{\dag}= -\frac{d^{2}}{dx^{2}}+V_{eff}(x),~~~~\mathcal{H}^{\dag}\Psi_1=\frac{E^{2}}{V^{2}_{F}}\Psi_1
\end{equation}
where
\begin{equation}\label{d30}
  \Psi_1(x)=\exp\left[\frac{1}{2}\int \left(2ieA_{x}(x)-a^{2}\sin x+\tan x \right)dx\right]\phi(x),
\end{equation}
\begin{equation}\label{d31}
  V_{eff}(x)=-\frac{V'^{2}_{F}}{4V^{2}_{F}}+\frac{V''_{F}}{2V_{F}}+\frac{(A_u(x)ae+k)^{2}}{R(x)^{2}}-\frac{aeA'_{u}(x)}{R(x)}+
 \frac{( k+aeA_{u}(x))R'(x)}{R(x)^{2}}-\frac{kV'_{F}}{R(x)V_{F}}-\frac{aeA_{u}(x)V'_{F}}{R(x)V_{F}},
\end{equation}
and $A_u(x)=a_{2}R(x)-\frac{k}{ae}$. Using $V_F(x)= a \cos x$, the effective potential $V_{eff}(x)$ becomes
\begin{equation}\label{d033}
  V_{eff}(x)=a^2a^{2}_{2}e^{2}-\frac{1}{2}+a_{2}e a\tan x-\frac{1}{4}\tan^{2} x,
\end{equation}
which is known as trigonometric Rosen-Morse-II potential in the literature \cite{sukhatme, Levai}. Let us highlight here that the ansatz on the Fermi velocity as a trigonometric cosine functions is a reasonable assumption due to the symmetry of the system.
\section{ Solutions}
\subsubsection{constant Fermi velocity: approximate solutions}
Our goal is to solve (\ref{d20}). First, let us consider the system below
\begin{equation}\label{d34}
  -\psi''(x)+(A+B \cos x+ C \sin x+D \sin^{2}x)\psi(x)=\varepsilon \psi(x),
\end{equation}
$\varepsilon$ is the eigenvalue and using a point transformation $z=\exp(ix)$, it becomes
\begin{equation}\label{d34}
  z^{2}\psi''(z)+z\psi'(z)+\left[A+\varepsilon+\frac{D}{2}+\frac{B-iC}{2}z+\frac{B+iC}{2}\frac{1}{z}-\frac{D}{4}\left(z^{2}+\frac{1}{z^{2}}\right)\right]\psi(x)=0.
\end{equation}
Expanding the coefficient of derivative-free term near $z=1$ up to the third order term gives
\begin{equation}\label{d35}
  z^{2}\psi''(z)+z\psi'(z)+(A+\varepsilon+B+iC(z-1)+\frac{1}{2}(B-iC-2D)(z-1)^{2})\psi(x)=0.
\end{equation}
Then, we apply the following transformation $z=\exp(-\alpha t)$ to get the equation given by
\begin{equation}\label{d36}
   \psi''(t)+\alpha^{2}\left[A+\frac{\varepsilon}{\alpha^2}+B+iC(\exp(-\alpha t)-1)+\frac{1}{2}(B-iC-2D)(\exp(-\alpha t)-1)^{2}\right]\psi(t)=0.~~~~-\infty < t < \infty
\end{equation}

For $C=0$, the potential is real and this also terminates the $\sin x$ function in the model. One can give the parameters of (\ref{d34}) in terms of original potential parameters given in (\ref{d20}) as
\begin{eqnarray}
  A &=& a^4(a^2+c^2)C^{2}_2 e^{2},~~~B=2ca^{5}C^{2}_{2}e^{2} \\
  C &=& eC_2 a^{2}(a-2),~~~~~D=-a^{6}C^{2}_{2}e^{2}.\label{d036}
\end{eqnarray}
Considering (\ref{d36}), for the real eigenvalues, $C$ should be pure imaginary $C=i\bar{C}$ which means that we shall take  $C_2$ as  $C_2 = i\bar{C_2}$ in (\ref{d036}) and (\ref{d36}). From (\ref{C2}), it can be seen that $a<1$. Now, the solutions of the model are already known in the literature \cite{yes, pr}. One can solve the eigenvalue equation below to get the real energies \cite{sukhatme}:
\begin{equation}\label{v}
  E_n=\pm \frac{V_{F} \alpha}{a} \sqrt{A+\frac{B+3\bar{C}}{2}-\left(\mu-n\right)^2},
\end{equation}
where $\mu=\frac{1}{2}\left(-1+\frac{\sqrt{2}(-B-2\bar{C}+2D)}{\sqrt{B+\bar{C}-2D}}\right)$, $A,B, \bar{C}$ are real parameters. And wavefunctions can be written as \cite{sukhatme}
\begin{equation}
\psi_{1,n}(z)\sim  z^{1-s}e^{-\frac{z}{2}}L^{2(s-n)}_{n}(z),
\end{equation}
with $s=\frac{1}{2\alpha}(-1+\frac{\sqrt{2}(B+2\bar{C}-2D)}{\sqrt{B+\bar{C}-2D}})$ and  $L_a^{c}(x)$ are the associated Laguerre polynomials. When $1-s >0$, the behaviours of the solutions are given in the limit of $z \rightarrow \infty$, $\psi_{1,n}(z) \rightarrow 0$, and $z \rightarrow 0$, $\psi_{1,n}(z) \rightarrow 0$. 
\subsubsection{position-dependent Fermi velocity }
$V_{eff}(x)$ in (\ref{d033}) is the element of the equation given below
\begin{equation}\label{d37}
  -\phi''(x)+V_{eff}(x)\phi(x)=\epsilon^{2}\phi(x).
\end{equation}
Using $\phi(x)=e^{-\frac{\alpha x}{2}} \phi_1(x)$, $z=\tan x$ and $\phi_{1}(z)=(1+z^{2})^{\beta}\bar{\phi}_{1}(z)$, we get
\begin{eqnarray}\label{d38}
  \nonumber(1+z^{2})\bar{\phi}_{1}(z)&+&(-\alpha+2z(1+2\beta))\bar{\phi}_{1}(z)+\\
  &&+\frac{1}{4(1+z^{2})}\left[(-4a_{2}e-8\alpha \beta)z+(1+4\beta)^{2}z^{2}-4C_1+\alpha^{2}+8\beta+4\epsilon^{2}\right]\bar{\phi}_{1}(z)=0.
\end{eqnarray}
For the values of the constant $a_2$ and $\beta$ as
\begin{eqnarray}
  a_{2} &=& -\frac{2\alpha \beta}{e} \\
  \beta &=& \frac{1}{4}\sqrt{-1-4C_1+\alpha^{2}+4\epsilon^{2}}.
\end{eqnarray}
(\ref{d38}) becomes
\begin{eqnarray}\label{d39}
  \nonumber(1+z^{2})\bar{\phi}^{''}_{1}(z)&+&\left(-\alpha+z(2+\sqrt{-1-4C_1+\alpha^{2}+4\epsilon^{2}})\right)\bar{\phi}^{'}_{1}(z)+\\
  &&+\left(\epsilon^{2}+\frac{\alpha^{2}-4C_1}{4}+
  \frac{1}{2}\sqrt{-1-4C_1+\alpha^{2}+4\epsilon^{2}}\right)\bar{\phi}_{1}(z)=0.
\end{eqnarray}
Now we can apply the new  variable as $s=\frac{1-iz}{2}$ and get
 \begin{eqnarray}\label{d40}
 \nonumber s(1-s)\frac{d^{2}\bar{\phi}_{1}(s)}{ds^{2}}&+&\left(-\frac{i\alpha}{2}+1+2\beta-2(1+2\beta)s\right)\frac{d\bar{\phi_{1}}}{ds}\\
 &&-\left(\epsilon^{2}+\frac{\alpha^{2}-4C_1}{4}+
  \frac{1}{2}\sqrt{-1-4C_1+\alpha^{2}+4\epsilon^{2}}\right)\bar{\phi}_{1}(s)=0.
\end{eqnarray}
(\ref{d40}) is the type of Hypergeometric differential equation in the literature \cite{abro}
\begin{equation}\label{d41}
  s(1-s)w''(s)+[\gamma-(a+b+1)s]w'(s)-abw(s)=0
\end{equation}
whose solutions are given by
\begin{equation}\label{d42}
  w(s)=c_1 ~~_{2}F_1(a,b;\gamma;s)+c_2s^{1-\gamma}~~_{2}F_1(a-\gamma+1,b-\gamma+1;2-\gamma;s).
\end{equation}
If we match (\ref{d41}) with (\ref{d40}), we get
\begin{eqnarray}
  \gamma &=& 1+2\beta-\frac{i\alpha}{2} \\
  a &=& \frac{1}{2}+2\beta+\frac{1}{2}\sqrt{5+16C_1-4\alpha^{2}+8\beta+16\beta^{2}-16\epsilon^{2}} \\
  b &=& \frac{1}{2}+2\beta-\frac{1}{2}\sqrt{5+16C_1-4\alpha^{2}+8\beta+16\beta^{2}-16\epsilon^{2}} 
\end{eqnarray}
Finally the solutions $ \phi (x)$ become polynomials if $a=-n$ or $b=-n$, then, the series terminates. The solutions have a form
\begin{equation}\label{d43}
  \phi(x)=N~\exp\left(-\frac{\alpha x}{2}\right) (1+\tan^{2} x)^{\beta}~~_{2}F_1\left(a,b;\gamma; \frac{1-i\tan x}{2}\right).
\end{equation}
If we look at the behaviour of the wavefunction, the hypergeometric function is defined through $|s| <1. $ When $s \rightarrow \pm 1$, ~~ $\phi(s)$ takes complex values. And the energy eigenvalues can be given by
\begin{equation}\label{d33}
   \epsilon_n=\pm \sqrt{ \frac{1}{2}(n+\mu+1)^{2}-\frac{1}{2}\frac{\nu}{(n+\mu+1)^{2}}}
\end{equation}
where $\mu$ and $\nu$ are the constants in terms of $\alpha, \beta, a_2$, ~$n=0,1,..$.

\section{Final Remarks}

In this paper we have studied the Dirac equation in $(2+1)$ dimensions  on the toroidal surface for a massless fermion particle under the action of external fields. Using the covariant approach based in general relativity, the Dirac operator stemming from a metric related to the strain tensor is discussed within the Pseudo-Hermitian operator theory.

We have initially obtained two coupled first-order differential equations coupling the left and right sector of the Dirac spinor. The decoupling of these equations renders two Klein-Gordon-like equations which were discussed in two cases, namely, constant and position-dependent Fermi velocity. 

The solution for both constant and position-dependent Fermi velocity cases were analytically obtained. In case of constant Fermi velocity calculations, we have obtained a condition on the inner radius $a < 1$  and we have extended the solutions of more general Mathieu potential whose solutions are given in terms of Laguerre polynomials. In the next case, the position-dependent Fermi velocity function is used to obtain the solutions in terms of hypergeometric functions with a trigonometric Rosen-Morse II type potential.

The paper not only presents important properties about the dynamics of an electron constrained to move on a torus surface under the action of external fields but also opens up new possibilities of investigation. The thermodynamic properties as well as electron-phonon interaction will be addressed in a future work.
\\
\\
\textbf{ DATA AVAILABILITY STATEMENT}
\\
The data that support the findings of this study are available from the corresponding author
upon reasonable request.

\end{document}